\documentclass[aps,prd,superscriptaddress,showpacs,preprint,amsmath,amssymb]{revtex4}
\usepackage{graphicx, bm}
\usepackage[usenames]{color}

\begin{document}

\draft
\title{Improved sensitivity on the electromagnetic dipole moments of the top quark in $\gamma\gamma$, $\gamma\gamma^*$ and
$\gamma^*\gamma^*$ collisions at the CLIC}

\author{A. A. Billur\footnote{abillur@cumhuriyet.edu.tr}}
\affiliation{\small Deparment of Physics, Cumhuriyet University, 58140, Sivas, Turkey.\\}

\author{M. K\"{o}ksal\footnote{mkoksal@cumhuriyet.edu.tr}}
\affiliation{\small Deparment of Optical Engineering, Cumhuriyet University, 58140, Sivas, Turkey.\\}

\author{ A. Guti\'errez-Rodr\'{\i}guez\footnote{alexgu@fisica.uaz.edu.mx}}
\affiliation{\small Facultad de F\'{\i}sica, Universidad Aut\'onoma de Zacatecas\\
         Apartado Postal C-580, 98060 Zacatecas, M\'exico.\\}

\date{\today}

\begin{abstract}

We realize a phenomenological study to examine the sensitivity on the magnetic moment and electric dipole moment of the top quark through the processes $\gamma\gamma\rightarrow t \bar{t}$, $e \gamma \rightarrow e \gamma^{*} \gamma \rightarrow e t \bar{t}$ and $e^{-} e^{+}\rightarrow e^{-} \gamma^{*} \gamma^{*} e^{+} \rightarrow e^{-} t \bar{t}e^{+}$ at the CLIC. We find that with a center-of-mass energy of the CLIC-1.4$\hspace{0.8mm}TeV$, integrated luminosity of ${\cal L}=1500\hspace{0.8mm}fb^{-1}$ and CLIC-3$\hspace{1mm}TeV$, integrated luminosity of ${\cal L}=2000\hspace{0.8mm}fb^{-1}$ with systematic uncertainties of $\delta_{sys}=0, 5, 10\hspace{1mm}\%$ at the $95\%\hspace{1mm}C. L.$, it is possible the CLIC may put limits on the electromagnetic dipole moments of the top quark $\hat a_V$ and $\hat a_A$ with a sensitivity of ${\cal O}(10^{-3}-10^{-2})$. Therefore, we show that the sensitivity with the CLIC data is much greater than that for the LHC data.
\end{abstract}


\vspace{5mm}

\maketitle

\section{Introduction}

The Standard Model (SM) has been tested in many important experiments and has been
quite successful, particularly after the discovery of a particle consistent with the Higgs
boson with a mass of about $125\pm 0.4 \hspace{0.8mm}GeV$. On the other hand, some of the most fundamental
questions still remain unanswered. For instance, the CP problem, neutrino oscillations
and matter-antimatter asymmetry have not been adequately clarified by the SM.
For this reason, it is often thought that the SM is embedded in a more fundamental theory with which its effects can
be observed at higher energy scales.

The top quark is the most massive of all observed elementary particles in the SM.
Because of the top quark's large mass, its couplings are expected to be more sensitive to new physics beyond the SM with respect to other particles. New physics can manifest itself in different forms. One possibility is that the new physics may lead to the appearance or a huge increase of new types of interactions like $tH^+b$ or anomalous Flavor Changing Neutral Current $tqg$, $tq\gamma$ and $tqZ$ ($q=u, c$) interactions. Another possibility is the modification of the SM couplings that involve $t\bar t g$, $t\bar t \gamma$, $t\bar t Z$ and $tW b$ vertices.

CP violation was first discovered in a small fraction of the kaon decays.
This phenomenology can be easily introduced by the Cabibbo-Kobayashi-Maskawa matrix (CKM) in the quark sector. CP violation in this sector is not enough
to clarify the baryon asymmetry in the universe. This asymmetry is one of the basic problems in the SM that has not been resolved even
in the heavy quarks decay processes. Therefore, the measurement of large amounts of CP violation in the top quark processes in colliders can demonstrate new physics. The existence of new physics can be analyzed by investigating the electromagnetic properties
of the top quark that are determined with its dipole moments such as the Magnetic Dipole Moment (MDM) and Electric Dipole Moment (EDM) defined as a source of CP violation.

The projection in the SM for the MDM of the top quark is $a^{SM}_t=0.02$ \cite{Benreuther},
and can be tested in the current and future colliders such as the Large Hadron Collider (LHC) and the Compact Linear Collider (CLIC). In contrast, the EDM of the top quark is strongly suppressed with a value of less than $10^{-30}\hspace{0.8mm}e  \hspace{1mm}{\mbox{cm}}$ \cite{Hoogeveen,Pospelov,Soni},
and is much too small to be observed. However, it is very attractive for probing new physics. If there is a new physics beyond the SM, the top quark may have a higher EDM value than $10^{-30}\hspace{0.8mm}e  \hspace{1mm}{\mbox{cm}}$. It is worth mentioning that the sensitivity to the
EDM has been studied in models with vector like multiplets which predicted the top quark EDM close to $1.75\times 10^{-3}$ \cite{Ibrahim}.

The studies performed through the $t\bar t\gamma$ production for the LHC at $\sqrt{s}=14\hspace{0.8mm}TeV$,
${\cal L}=300\hspace{0.8mm}fb^{-1}$ and $3000\hspace{0.8mm}fb^{-1}$ reported the limits of $\pm 0.2$ and $\pm 0.1$,
respectively \cite{Baur}. The limits $-2.0\leq \hat a_V\leq 0.3$ and $-0.5\leq \hat a_A\leq 1.5$
are obtained from the branching ratio and the CP asymmetry from radiative $b \to s\gamma$ transitions \cite{Bouzas}.
However, the authors of Ref. \cite{Bouzas1} obtained the bounds on $|\hat a_V|< 0.05\hspace{0.8mm}(0.09)$
and $|\hat a_A|< 0.20\hspace{0.8mm}(0.28)$ from measurements of the $\gamma^{*} p\to t\bar t$ cross
section with $10\%$ $(18\%)$ uncertainty at the Large Hadron electron Collider (LHeC), respectively. Bounds on the dipole moments
of the top quark were recently reported in literature through the process $pp \to p\gamma^* \gamma^* p \to p t\bar tp$ for the
energy and luminosity of the LHC of $\sqrt{s}=14\hspace{0.8mm}TeV$, ${\cal L}=3000\hspace{0.8mm}fb^{-1}$ and $68\%$ C.L.:
$-0.6389\leq \hat a_V\leq 0.0233$ and $|\hat a_A|\leq 0.1158$ \cite{Sh}.

Moreover, in the case of the $e^+e^-$ collider as the International Linear Collider (ILC), the sensitivity bounds at $1\sigma$ for the anomalous couplings of the top quark through top quark pair production $e^+e^- \to t\bar t$ at $\sqrt{s}=500\hspace{0.8mm}GeV$, ${\cal L}=200\hspace{0.8mm}fb^{-1}$, ${\cal L}=300\hspace{0.8mm}fb^{-1}$ and ${\cal L}=500\hspace{0.8mm}fb^{-1}$ are predicted to be of the order ${\cal O}(10^{-3})$, indicating that measurements at an electron positron collider lead to a significant improvement in comparison with the LHC.
Thorough and detailed discussions on the dipole moments of the top quark in top quark pair production at the ILC are reported in
the literature \cite{Atwood,Polouse,Choi,Polouse1,Aguilar0,Amjad,Juste,Asner,Abe,Aarons,Brau,Baer}. On the other hand,  Ref. \cite{Grzadkowski:2005ye} have found that the process $e^{-}e^{+} \rightarrow t\bar{t}$  will do slightly better than $\gamma \gamma\rightarrow t\bar{t}$ for the determination of the anomalous $tt \gamma$ couplings.

In Ref. \cite{murat}, bounds are estimated on the electromagnetic dipole moments of the top quark through the processes $\gamma e^- \to \bar t b\nu_e$
and $e^+e^- \to e^-\gamma^* e^+ \to \bar t b\nu_e e^+$ with unpolarized and polarized electron beams at the CLIC. For the systematic uncertainties of $\delta_{sys}=0\%,\hspace{1mm}5\%$, $b-\mbox{tagging efficiency}=0.8$, center-of-mass
energy of $\sqrt{s}=3\hspace{0.8mm}TeV$, integrated luminosity of ${\cal L}=2000\hspace{0.8mm}fb^{-1}$ and $2\sigma\hspace{1mm}(3\sigma)$,
the bounds obtained on the electromagnetic dipole moments $\hat a_V$ and $\hat a_A$ of the top quark are of the order ${\cal O}(10^{-2}-10^{-1})$
and are highly competitive with those reported in previous studies.

The advantage of the linear $e^{-}e^{+}$ colliders with respect to the hadron colliders is in the general cleanliness of the events where two elementary particles, electrons and positrons beams, collide at high energy, and the high resolutions of the detector made possible by the relatively low absolute rate of background events. In addition, these colliders will complement the physics program of the LHC, especially for precision physics. Therefore, precise measurements of the top quark properties, such as the mass, charge, spin and
dipole moments will become possible. The CLIC is a proposed future $e^{-}e^{+}$ collider, designed to
fulfill $e^{-}e^{+}$ collisions at center-of-mass energies of 0.35, 1.4 and 3 TeV planned to be constructed
with a three main stage research region \cite{Abramowicz}. This enables the investigation of the $\gamma\gamma$ and $e\gamma$
interactions by converting the original $e^{-}$ or $e^{+}$ beam into a photon beam through the
Compton backscattering mechanism. The other well-known applications of the linear colliders are the processes $e\gamma^{*}$, $\gamma \gamma^{*}$ $\gamma^{*} \gamma^{*}$ where the emitted quasireal photon $\gamma^{*}$ is scattered with small angles from the beam pipe of $e^{-}$ or $e^{+}$ \cite{Ginzburg,Ginzburg1,Brodsky,Budnev,Terazawa,Yang}. Since these photons have a low virtuality, they are almost on the mass shell. These processes can be described by the Weizsacker-Williams Approximation (WWA). The WWA has a lot of advantages such as providing the skill to reach crude numerical predictions via simple formulae. In addition, it may principally ease the experimental analysis because it enables one to directly achieve a rough cross section for $\gamma^{*} \gamma^{*}\rightarrow X$ process via the examination of the main process $e^{-}e^{+}\rightarrow e^{-} X e^{+}$ where X represents objects produced in the final state. The production of high mass objects is particularly interesting at the linear colliders and the production rate of massive objects is limited by the photon luminosity at high invariant mass while $\gamma^{*}\gamma^{*}$ and $e\gamma^{*}$ processes at the linear colliders arise from quasireal photon emitted from the incoming beams. Hence, $\gamma^{*} \gamma^{*}$ and $e\gamma^{*}$ are more realistic than $\gamma\gamma$ and $e\gamma$. These processes have been observed experimentally at LEP, Tevatron and LHC \cite{Abulencia,Aaltonen1,Aaltonen2,Chatrchyan1,Chatrchyan2,Abazov,Chatrchyan3,Inan,Inan1,Inan2,Sahin1,Atag,Sahin2,Sahin4,Senol,Senol1,Fichet,Sun,Sun1,Sun2,Senol2,Atag1,Koksal1,koksal2,koksal3,koksal4,billur1, koksal5,koksal6,arı1,koksal7,billur2,billur3,Fichet1,Sun3}.

In this paper, we perform a phenomenological study for determining the sensitivity on the magnetic moment and electric dipole moment
of the top quark through the $t\bar t$ pair production in $e^+e^-$ colliders, specifically for center-of-mass energy and luminosity
of CLIC-1.4$\hspace{0.8mm}TeV$, ${\cal L}=1500\hspace{0.8mm}bf^{-1}$ and CLIC-3$\hspace{0.8mm}TeV$, ${\cal L}=2000\hspace{0.8mm}bf^{-1}$
with systematic uncertainty of $\delta_{sys}=0, 5, 10\hspace{0.8mm}\%$ and $95\%\hspace{0.8mm}C.L.$.
Here, we consider that the top quark pair production in $e^+e^-$ interactions are given through three different processes $\gamma\gamma\rightarrow t \bar{t}$, $e \gamma \rightarrow e \gamma^{*} \gamma \rightarrow e t \bar{t}$, $e^{-} e^{+}\rightarrow e^{-} \gamma^{*} \gamma^{*} e^{+} \rightarrow e^{-} t \bar{t}e^{+}$ where $\gamma$ and $\gamma^{*}$ are Compton backscattered and Weizsacker-Williams photons, respectively. These processes are one of the most important sources of $t\bar t$ pair production and may represent new physics effects
at a high energy and high luminosity linear electron positron collider such as the CLIC and also isolate anomalous $t\bar t\gamma$ coupling from $t\bar t Z$.

This work is structured as follows. In Section II, we introduce the top quark effective electromagnetic interactions. In Section III,
we study the dipole moments of the top quark through the processes $\gamma\gamma\rightarrow t \bar{t}$, $e \gamma \rightarrow e \gamma^{*} \gamma \rightarrow e t \bar{t}$ and $e^{-} e^{+}\rightarrow e^{-} \gamma^{*} \gamma^{*} e^{+} \rightarrow e^{-} t \bar{t}e^{+}$. Finally, we present our conclusions in Section IV.

\section{Top quark pair production processes in photon-photon collisions}

\subsection{General Effective Coupling $t\bar t \gamma$}

The most general effective coupling $t\bar t\gamma$ which includes the SM coupling and contributions from dimension-six
effective operators can be written as \cite{Sh,Kamenik,Baur,Aguilar,Aguilar1}:

\begin{equation}
{\cal L}_{t\bar t\gamma}=-g_eQ_t\bar t \Gamma^\mu_{ t\bar t  \gamma} t A_\mu,
\end{equation}

\noindent where $g_e$ is the electromagnetic coupling constant, $Q_t$ is the top quark electric charge
and the Lorentz-invariant vertex function $\Gamma^\mu_{t\bar t \gamma}$, which describes the
interaction of a $\gamma$ photon with two top quarks, can be parameterized by

\begin{equation}
\Gamma^\mu_{t\bar t\gamma}= \gamma^\mu + \frac{i}{2m_t}(\hat a_V + i\hat a_A\gamma_5)\sigma^{\mu\nu}q_\nu,
\end{equation}

\noindent where $m_t$ is the mass of the top quark, $q$ is the momentum transfer to the photon and the couplings
$\hat a_V$ and $\hat a_A$ are real and related to the anomalous magnetic moment and the electric dipole moment
of the top quark, respectively.

\subsection{Theoretical Calculations}

\begin{figure}[t]
\centerline{\scalebox{0.8}{\includegraphics{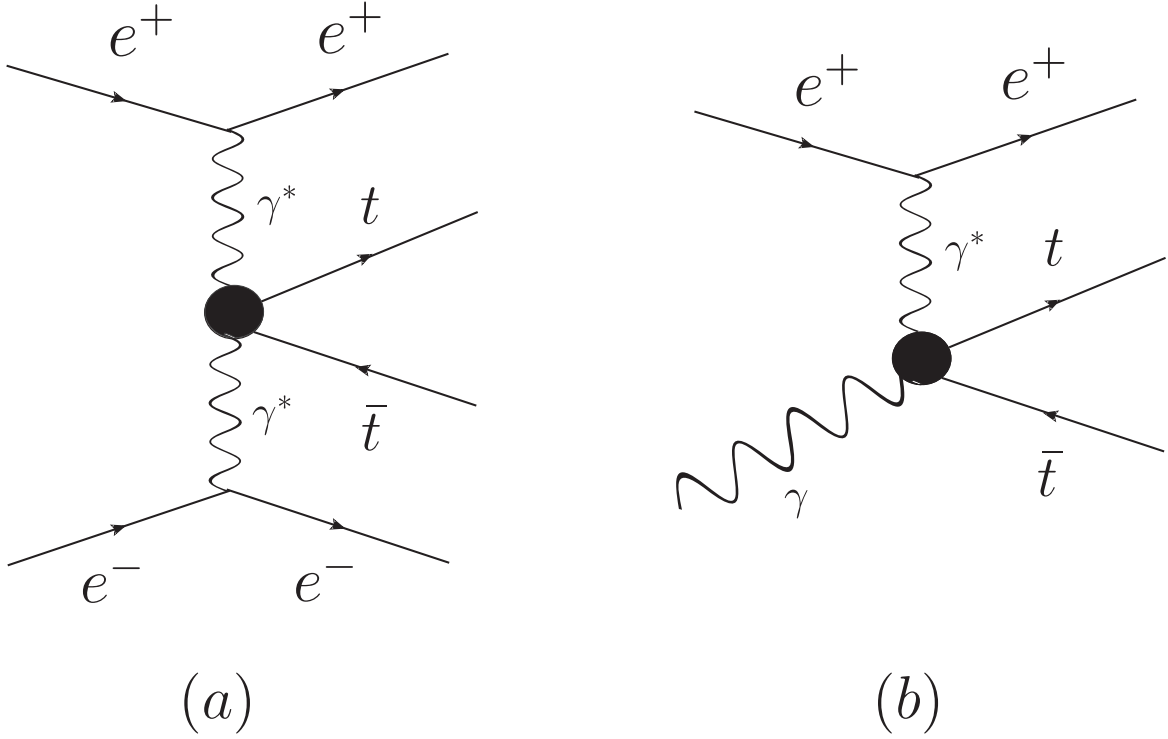}}}
\caption{ \label{fig:gamma1} A schematic diagram for the process (a) $e^{-} e^{+}\rightarrow e^{-} \gamma^{*} \gamma^{*} e^{+} \rightarrow e^{-} t \bar{t}e^{+}$  and (b) $e^{+} \gamma \rightarrow e^{+} \gamma^{*} \gamma \rightarrow e^{+} t \bar{t}$.}
\label{Fig.1}
\end{figure}

\begin{figure}[t]
\centerline{\scalebox{0.8}{\includegraphics{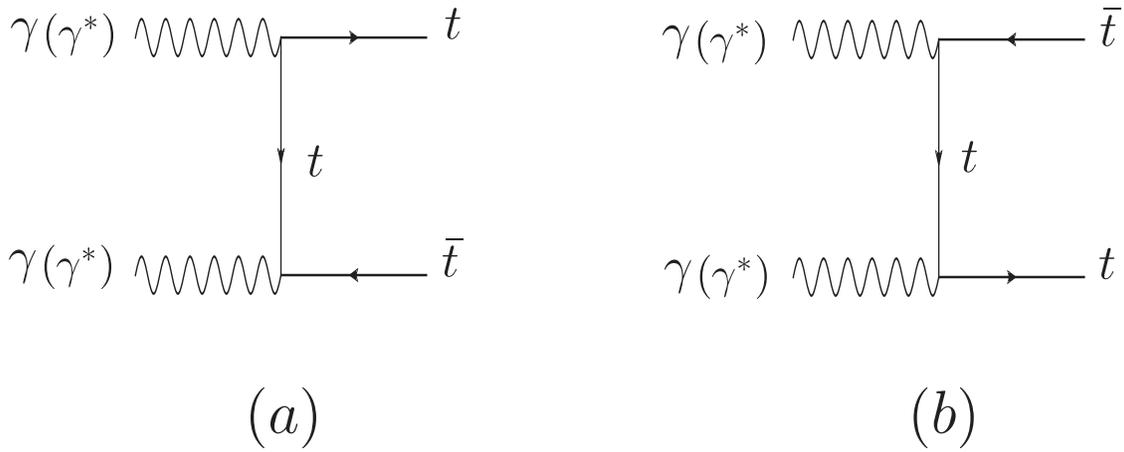}}}
\caption{ \label{fig:gamma2} Feynman diagrams contributing to the process
$\gamma\gamma \to t \bar t$ and the subprocesses $\gamma\gamma^* \to t \bar t$ and $\gamma^*\gamma^* \to t \bar t$.}
\label{Fig.2}
\end{figure}

Schematic diagrams for the processes $e \gamma \rightarrow e \gamma^{*} \gamma \rightarrow e t \bar{t}$ and $e^{-} e^{+}\rightarrow e^{-} \gamma^{*} \gamma^{*} e^{+} \rightarrow e^{-} t \bar{t}e^{+}$ are given in Fig. \ref{Fig.1}. With these processes, $ \gamma (\gamma^*) \gamma(\gamma^*) \rightarrow t \bar{t} $ have two Feynman diagrams which are shown in detail in Fig. \ref{Fig.2}.

For $\gamma\gamma$, $\gamma\gamma^{*}$ and $\gamma^{*}\gamma^{*}$ collisions including the effective Lagrangian in Eq. 1, the polarization
summed amplitude square is given in function of the Mandelstam invariants $\hat{s}$, $\hat{t}$
and $\hat{u}$ as follows,

\begin{eqnarray}
|M_{1}|^{2}&=&\frac{16\pi^{2}Q_{t}^2\alpha^{2}_e}{2m_{t}^{4}(\hat{t}-m_{t}^{2})^{2}}\biggl[48\hat{a}_{V}(m_{t}^{2}-\hat{t})
(m_{t}^{2}+\hat{s}-\hat{t})m_{t}^{4}-16(3m_{t}^{4}-m_{t}^{2}\hat{s}+\hat{t}(\hat{s}+\hat{t})) m_{t}^{4}\nonumber\\
&+&2(m_{t}^{2}-\hat{t})(\hat{a}_{V}^{2}(17m_{t}^{4}+(22\hat{s}-26\hat{t})m_{t}^{2} +\hat{t}(9\hat{t}-4\hat{s}))  \nonumber\\
&+&\hat{a}_{A}^{2}(17m_{t}^{2}+4\hat{s}-9\hat{t})(m_{t}^{2}-\hat{t}))m_{t}^{2}+12\hat{a}_{V}(\hat{a}_{V}^{2}+\hat{a}_{A}^{2})\hat{s}(m_{t}^{3}-m_{t}\hat{t})^{2}\nonumber\\
&-&(\hat{a}_{V}^{2}+\hat{a}_{A}^{2})^{2}(m_{t}^{2}-\hat{t})^{3}(m_{t}^{2}-\hat{s}-\hat{t})\biggr],
\end{eqnarray}

\begin{eqnarray}
|M_{2}|^{2}&=&\frac{-16\pi^{2}Q_{t}^2\alpha^{2}_e}{2m_{t}^{4}(\hat{u}-m_{t}^{2})^{2}}\biggl[48\hat{a}_{V}(m_{t}^{4}+(\hat{s}-2\hat{t})m_{t}^{2}+\hat{t}(\hat{s}+\hat{t}))m_{t}^{4}\nonumber\\
&+&16(7m_{t}^{4}-(3\hat{s}+4\hat{t})m_{t}^{2}+\hat{t}(\hat{s}+\hat{t})) m_{t}^{4}\nonumber\\
&+&2(m_{t}^{2}-\hat{t})(\hat{a}_{V}^{2}(m_{t}^{4}+(17\hat{s}-10\hat{t})m_{t}^{2}+9\hat{t}(\hat{s}+\hat{t})) \nonumber\\
&+&\hat{a}_{A}^{2}(m_{t}^{2}-9\hat{t})(m_{t}^{2}-\hat{t}-\hat{s}))m_{t}^{2}\nonumber\\
&+&(\hat{a}_{V}^{2}+\hat{a}_{A}^{2})^{2}(m_{t}^{2}-\hat{t})^{3}(m_{t}^{2}-\hat{s}-\hat{t})\biggr],
\end{eqnarray}

\begin{eqnarray}
M_{1}^{\dag}M_{2}+M_{2}^{\dag}M_{1}&=&\frac{16\pi^{2}Q_{t}^2\alpha^{2}_e}{m_{t}^{2}(\hat{t}-m_{t}^{2})(\hat{u}-m_{t}^{2})} \nonumber \\
&\times &\biggl[-16(4m_{t}^{6}-m_{t}^{4}\hat{s})+8\hat{a}_{V}m_{t}^{2}(6m_{t}^{4}-6m_{t}^{2}(\hat{s}+2\hat{t})-\hat{s})^{2} \nonumber \\ &+&6\hat{t})^{2}+6\hat{s}\hat{t})+(\hat{a}_{V}^{2}(16m_{t}^{6}-m_{t}^{4}(15\hat{s}+32\hat{t})+m_{t}^{2}(15\hat{s})^{2} \nonumber \\
&+&14\hat{t}\hat{s}+16\hat{t})^{2})+\hat{s}\hat{t}(\hat{s}+\hat{t}))+\hat{a}_{A}^{2}(16m_{t}^{6}-m_{t}^{4}(15\hat{s}+32\hat{t})  \nonumber\\
&+&m_{t}^{2}(5\hat{s})^{2}+14\hat{t}\hat{s}+16\hat{t})^{2})+\hat{s}\hat{t}(\hat{s}+\hat{t})))-4\hat{a}_{V}\hat{s}(\hat{a}_{V}^{2}+\hat{a}_{A}^{2})\nonumber\\
&\times& (m_{t}^{4}+m_{t}^{2}(\hat{s}-2\hat{t})+\hat{t}(\hat{s}+\hat{t}))-4\hat{a}_{A}(\hat{a}_{V}^{2}+\hat{a}_{A}^{2})(2m_{t}^{2}
-\hat{s}-2\hat{t}) \nonumber\\
&\times& \epsilon_{\alpha \beta \gamma \delta} p_{1}^{\alpha}p_{2}^{\beta}p_{3}^{\gamma}p_{4}^{\delta}-2\hat{s}(\hat{a}_{V}^{2}+\hat{a}_{A}^{2})^{2}
(m_{t}^{4}-2\hat{t}m_{t}^{2}+\hat{t}(\hat{s}+\hat{t}))\biggr],
\end{eqnarray}

\noindent where $\hat s=(p_1 + p_2)^2=(p_3 + p_4)^2$, $\hat t=(p_1 - p_3)^2=(p_4 - p_2)^2$, $\hat u=(p_3 - p_2)^2=(p_1 - p_4)^2$ and
$p_{1}$ and $p_{2}$ are the four-momenta of the incoming photons, $p_{3}$ and $p_{4}$ are the momenta of the outgoing top quarks,
$Q_{t}$ is the top quark charge, $\alpha_e=g^2_e/4\pi$ is the fine-structure constant, $m_t$ is the mass of top and $\hat a_V$ ($\hat a_A$)
are their dipole moments.

The most promising mechanism to generate energetic photon beams in a linear collider is Compton backscattering. Compton backscattered
photons interact with each other and generate the process $\gamma \gamma \rightarrow t \bar{t}$. The spectrum of Compton backscattered
photons is given by

 \begin{eqnarray}
 f_{\gamma}(y)=\frac{1}{g(\zeta)}[1-y+\frac{1}{1-y}-
 \frac{4y}{\zeta(1-y)}+\frac{4y^{2}}{\zeta^{2}(1-y)^{2}}] ,
 \end{eqnarray}

 where

 \begin{eqnarray}
 g(\zeta)=(1-\frac{4}{\zeta}-\frac{8}{\zeta^2})\log{(\zeta+1)}+
 \frac{1}{2}+\frac{8}{\zeta}-\frac{1}{2(\zeta+1)^2} ,
 \end{eqnarray}

 with

 \begin{eqnarray}
 y=\frac{E_{\gamma}}{E_{e}} , \;\;\;\; \zeta=\frac{4E_{0}E_{e}}{M_{e}^2}
 ,\;\;\;\; y_{max}=\frac{\zeta}{1+\zeta}.
 \end{eqnarray}

Here, $E_{0}$ and $E_{e}$ are  energy of the incoming laser photon and initial energy of the electron beam before
Compton backscattering and $E_{\gamma}$ is the energy of the backscattered photon. The maximum value of $y$ reaches 0.83 when $\zeta=4.8$.

WWA is  another possibility for top pair production. The quasireal photons emitted from both lepton beams collide
with each other and produce the process $\gamma^{*} \gamma^{*} \rightarrow t \bar{t}$. In WWA, the photon spectrum is given by

\begin{eqnarray}
f_{\gamma^{*}}(x)=\dfrac{\alpha}{\pi E_{e}}\lbrace [\dfrac{1-x+x^2/2}{x}]log(\dfrac{Q^2_{max}}{Q^2_{min}})-\dfrac{m_{e}^2x}{Q^2_{min}}(1-\dfrac{Q^2_{min}}{Q^2_{max}})-\dfrac{1}{x}[1-\dfrac{x}{2}]^2log(\dfrac{x^2 E_{e}^2+Q^2_{max}}{x^2 E_{e}^2+Q^2_{min}})\rbrace,
\end{eqnarray}

\noindent where $x=E_{\gamma}/E_{e}$ and $Q^2_{max}$ is maximum virtuality of the photon. In this work, we have taken into account the maximum  virtuality of the photon is $Q^2_{max}=2\hspace{0.8mm}GeV^2$. The larger values of $Q^2_{max}$  do not make a significant contribution to the sensitivity limits which is similar to results in previous works \cite{alp,Gutierrez,Gutierrez1,atagx}. The minimum value of the $Q^2_{min}$ is given by

\begin{eqnarray}
Q^2_{min}=\dfrac{m_{e}^2x^2}{1-x}.
\end{eqnarray}

The $Q^2_{min}$ value is very small due to the electron mass. However, the scattering angles of the electrons are so small that the transverse momentum is close to zero. Due to the momentum conservation, the transverse momentum of the emitted photons also have small values. In light of all these arguments, virtuality of the photons in WWA is very small and the photons are almost on mass shell.

The process $\gamma^{*} \gamma^{*} \rightarrow t \bar{t}$ participates
as a subprocess in the main process $e^{-} e^{+}\rightarrow e^{-} \gamma^{*} \gamma^{*} e^{+} \rightarrow e^{-} t \bar{t}e^{+}$. However, an $\gamma^{*}$ photon emitted from either of the incoming leptons can interact with the Compton backscattered photon and the subprocess
$\gamma \gamma^{*} \rightarrow t \bar{t}$ can take place. Hence, we calculate the process $e \gamma \rightarrow e \gamma^{*} \gamma \rightarrow e t \bar{t}$ by integrating the cross section for the subprocess $\gamma \gamma^{*} \rightarrow t \bar{t}$.

The total cross sections are,
\begin{eqnarray}
\sigma=\int f_{\gamma(\gamma^{*})}(x)f_{\gamma(\gamma^{*})}(x)d\hat{\sigma}dE_{1}dE_{2}.
\end{eqnarray}

\begin{figure}[t]
\centerline{\scalebox{0.75}{\includegraphics{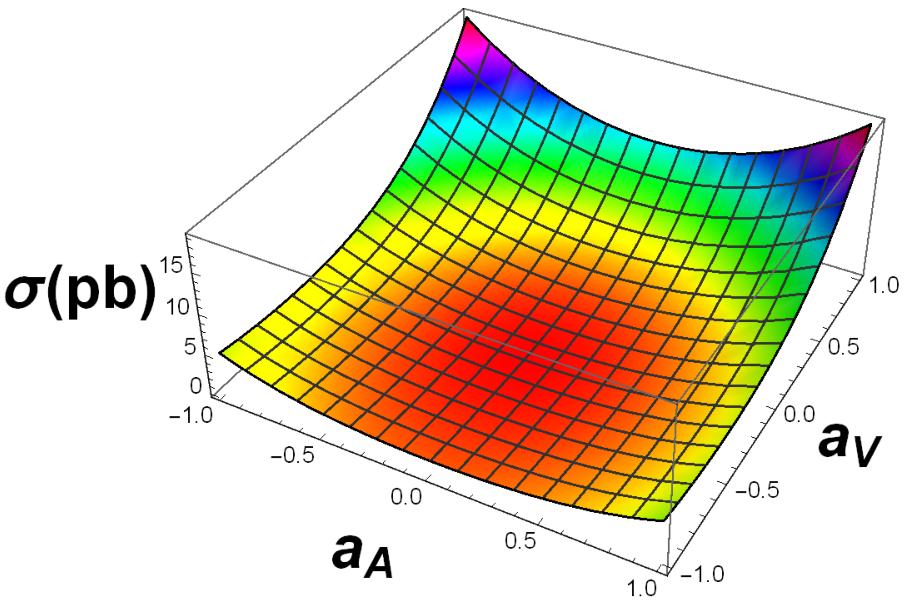}}}
\caption{ The total cross sections of the process
$\gamma \gamma \to t \bar t $ as a function of $\hat a_V$
and $\hat a_A$ for center-of-mass energy of
$\sqrt{s}=1.4$\hspace{0.8mm}$TeV$.}
\label{Fig.3}
\end{figure}

\begin{figure}[t]
\centerline{\scalebox{0.75}{\includegraphics{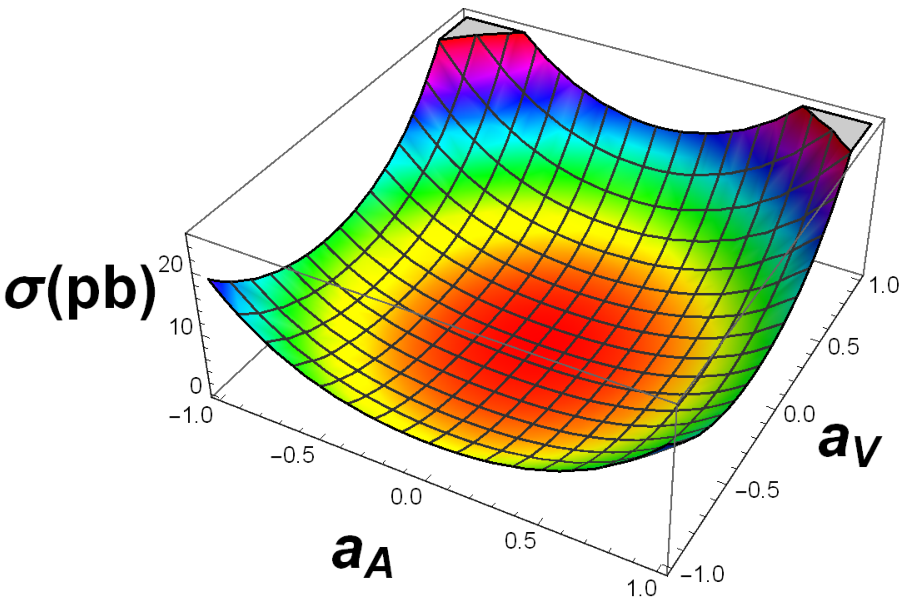}}}
\caption{ Same as in Fig. 3, but for center-of-mass energy of
$\sqrt{s}=3$\hspace{0.8mm}$TeV$.}
\label{Fig.4}
\end{figure}

\begin{figure}[t]
\centerline{\scalebox{0.75}{\includegraphics{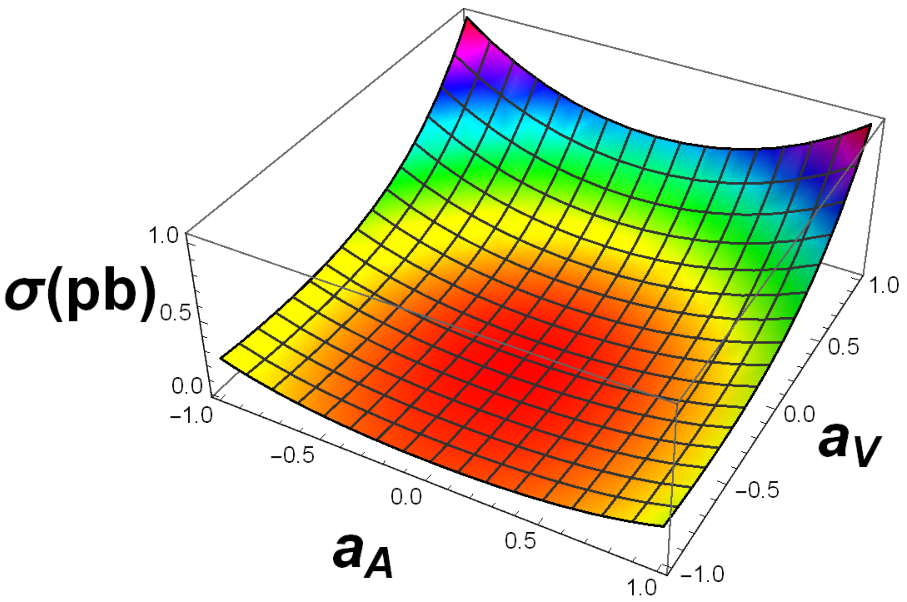}}}
\caption{ \label{fig:gamma1} The total cross sections of the process
$e^{+} \gamma \rightarrow e^{+} \gamma^{*} \gamma \rightarrow e^{+} t \bar{t}$ as a function of $\hat a_V$ and $\hat a_A$ for center-of-mass energy of
$\sqrt{s}=1.4$\hspace{0.8mm}$TeV$.}
\label{Fig.5}
\end{figure}

The total cross sections of these processes as functions of anomalous $\hat a_V$ and $\hat a_A$ are shown in Figs. \ref{Fig.3}-\ref{Fig.5}. In these figures, the cross sections depending on the anomalous couplings were obtained by varying only one of the anomalous couplings at a time while
the other was fixed to zero. We understand from Figs. \ref{Fig.3}-\ref{Fig.5} that the total cross sections show a clear dependence on the dipole moments of the top quark. Anomalous $\hat a_V$ and $\hat a_A$ parameters have different CP properties which can be seen in Eqs. 3-5. The cross section with respect to the $\hat a_A$ parameter is even power and a nonzero value of this parameter allows a constructive effect on the total cross section. In addition, the contribution of $\hat a_V$ coupling to the cross sections is proportional to odd power. In Fig. \ref{Fig.3}, there are small intervals around $\hat a_V$ in which the cross section that includes new physics is smaller than the SM cross section. For this reason, the $\hat a_V$ parameter has a partially destructive effect on the total cross section.

The scattering amplitudes can be given in Eqs. (3)-(5) as a polynomial in powers of $\hat a_V ( \hat a_A )$. Therefore, the cross section as a polynomial in powers of $\hat a_V$$ ( \hat a_A )$ for the three modes $\gamma\gamma \to t\bar t$, $e^+\gamma \to e^+\gamma^* \gamma \to e^+ t \bar t$ and $e^+e^- \to e^+\gamma^* \gamma^* e^- \to e^+ t \bar t e^-$
are given by

\begin{eqnarray}
\sigma_{Tot}(\hat a_V)&=&\sigma_4 \hat a^4_V + \sigma_3 \hat a^3_V + \sigma_2 \hat a^2_V + \sigma_1 \hat a^1_V + \sigma_0,     \\
\sigma_{Tot}(\hat a_A)&=&\sigma'_4 \hat a^4_A + \sigma'_2 \hat a^2_A + \sigma_0,
\end{eqnarray}

\noindent where $\sigma^i (\sigma^{'i})$ $i=1,..,4$ is the anomalous contribution, while $\sigma_0$ is the contribution of the
SM at $\hat a_V= \hat a_A =0$, respectively. This provides more precise and convenient information for each process. The numerical computations of the coefficients of $\hat a_V$  and $\hat a_A $ of the Eqs. (12) and (13) are presented in Table I. 

\begin{table}[!ht]
\caption{Numerical computations of the total cross sections versus $\hat a_V$  and $\hat a_A $ at $\sqrt{s}=$3 TeV.}
\begin{center}
\begin{tabular}{|c| c| c| c| c|| c| c| c| c|}
\hline
Mode              &  $\sigma_4$  & $\sigma_3$  & $\sigma_2$  & $\sigma_1$ & $\sigma_0$  &  $\sigma'_4$  & $\sigma'_2$   \\
\hline
\hline
$\gamma\gamma \to t\bar t$    &     $4.52$      &    $4.51$     &    $5.24$       &     $0.97$       &      $0.38$       &     $4.52$       &  $4.75$                                  \\
\hline
$e^+\gamma \to e^+\gamma^* \gamma \to e^+ t \bar t$  &     $0.24$       &    $0.42$     &    $0.56$     &       $0.19$      &     $0.07$     &    $0.24$      &       $0.46$           \\
\hline
$e^+e^- \to e^+\gamma^* \gamma^* e^- \to e^+ t \bar t e^-$        &   0.012      &   0.027       &    0.039     &   0.016     &     0.006     &    0.012      &     0.031                \\
\hline\hline
\end{tabular}
\end{center}
\end{table}

When comparing the three processes in Figs. \ref{Fig.3}-\ref{Fig.5}, the largest deviation from the SM of the anomalous cross
sections, including anomalous $\hat a_V$ and $\hat a_A$ couplings, is the process $\gamma\gamma \to t \bar t$. 
The best sensitivities on anomalous $\hat a_V$ and $\hat a_A$ couplings are obtained from the 
process $\gamma\gamma \to t \bar t$. Similarly, the sensitivities obtained on anomalous couplings through the 
process $e \gamma \rightarrow e \gamma^{*} \gamma \rightarrow e t \bar{t}$ are expected to be more restrictive 
than the sensitivities on the process 
$e^{-} e^{+}\rightarrow e^{-} \gamma^{*} \gamma^{*} e^{+} \rightarrow e^{-} t \bar{t}e^{+}$.

When making a direct comparison of our results for the total cross section as a function of the dipole moments $\hat a_V$ and $\hat a_A$ reported in Figs. \ref{Fig.3}-\ref{Fig.5} with those reported in Ref. \cite{Sh} (see Figs. \ref{Fig.3}-\ref{Fig.4}), we find that our results, using processes
$\gamma\gamma \to t \bar t$, $\gamma\gamma^* \to t \bar t$ and $\gamma^*\gamma^* \to t \bar t$ at CLIC energies, with respect to process $pp\to p\gamma^* \gamma^* p \to pt\bar t p$ at LHC energies, show a significant improvement. In addition, with our processes the total cross sections are of 3-4 orders of magnitude better than those reported in Ref. \cite{Sh}. This shows that the bounds on the anomalous couplings $\hat a_V$ and $\hat a_A$ can be improved at a linear collider such as the CLIC by a few orders of magnitude when compared to what is possible at the LHC.

\section{Dipole moments of the top quark in $\gamma\gamma$, $\gamma\gamma^*$ and $\gamma^*\gamma^*$ collisions}

To investigate the sensitivity to the anomalous $\hat a_V$ and $\hat a_A$ couplings we use the chi-squared distribution:

\begin{equation}
\chi^2=\biggl(\frac{\sigma_{SM}-\sigma_{NP}(\hat a_V, \hat a_A)}{\sigma_{SM}\delta}\biggr)^2,
\end{equation}

\noindent where $\sigma_{NP}(\hat a_V, \hat a_A)$ is the total cross section including contributions from the SM
and New physics, $\delta=\sqrt{(\delta_{st})^2+(\delta_{sys})^2}$, $\delta_{st}=\frac{1}{\sqrt{N_{SM}}}$
is the statistical error and $\delta_{sys}$ is the systematic error. The number of events for each of the three processes is given by
$N_{SM}={\cal L}_{int}\times BR \times \sigma_{SM}\times \epsilon_{b-tag}\times \epsilon_{b-tag}$, where ${\cal L}_{int}$ is the integrated luminosity and  b-jet tagging efficiency is 0.8 \cite{atlas}. The top quarks decay almost 100$\%$ to $W$ boson and b quark. For top quark pair production we can categorize decay products according to the decomposition of $W$. In this work, we assume that one of the $W$ bosons decays leptonically and the other hadronically for the signal. This phenomenon has already been studied by ATLAS and CMS Collaborations \cite{cmstop1,cmstop2,atlastop}. Thus, we assume that the branching ratio of the top quark pair in the final state to be BR = 0.286.

For our numerical computation, we take a set of independent input parameters which are known from current experiments. The input parameters are $\alpha=\frac{1}{137.4}$, $m_b=4.18\hspace{0.8mm}GeV$,
$m_t=173.21\hspace{0.8mm}GeV$ \cite{Data2014} and for our analysis, we consider a $95\%$ C.L. sensitivity on the dipole moments $\hat a_V$ and $\hat a_A$ of the top quark via the processes $\gamma\gamma\rightarrow t \bar{t}$, $e \gamma \rightarrow e \gamma^{*} \gamma \rightarrow e t \bar{t}$ and $e^{-} e^{+}\rightarrow e^{-} \gamma^{*} \gamma^{*} e^{+} \rightarrow e^{-} t \bar{t}e^{+}$ at the CLIC-1.4 $TeV$ with ${\cal L}_{int}=1500\hspace{0.8mm}fb^{-1}$ and CLIC-3 $TeV$ with ${\cal L}_{int}=2000\hspace{0.8mm}fb^{-1}$.

Tables I-VI illustrate the sensitivity obtained at $95\%$ $C.L.$ on the dipole moments $\hat a_V$ and $\hat a_A$
of the top quark through the processes $\gamma\gamma\rightarrow t \bar{t}$, $e \gamma \rightarrow e \gamma^{*} \gamma \rightarrow e t \bar{t}$ and $e^{-} e^{+}\rightarrow e^{-} \gamma^{*} \gamma^{*} e^{+} \rightarrow e^{-} t \bar{t}e^{+}$. The bounds are obtained assuming that the center-of-mass energy of
CLIC-1.4$\hspace{0.8mm}TeV$ and luminosities of ${\cal L}=50, 300, 500, 1000, 1500\hspace{0.8mm}fb^{-1}$ for the second
stage of operation of the collider. For the third stage, we consider the center-of-mass energy
of CLIC-3$\hspace{0.8mm}TeV$ and luminosities of ${\cal L}=50, 300, 500, 1000, 1500, 2000\hspace{0.8mm}fb^{-1}$.

An important part of our analysis is the inclusion of theoretical uncertainties as there may be several experimental and systematic uncertainty sources in top quark identification. This situation has not been studied experimentally in linear colliders. For hadron colliders, especially LHC, the process of determining the cross section of top pair production has been experimentally studied \cite{topuncertanintyatlas,topuncertaintycms}. As seen from these studies, the total systematic uncertainty value is about 10$\%$ and is increasingly improved when it is compared with previous experimental studies \cite{cmstop2}.

We use three scenarios for the systematic uncertainties in our entire set of
Tables: (1) we assume a systematic uncertainty of $\delta_{sys} = 0\%$, (2) we
estimate future results for $\hat a_V$ and $\hat a_A$ with $5\%$ systematic uncertainty and (3) we consider a systematic uncertainty
of as much as $\delta_{sys} = 10\%$. We find in Tables I-VI that the most prominent mode of top quark pair production that imposes stronger limits on the dipole moments is the production process
$\gamma\gamma \to t \bar t$, followed in order of importance by the processes $e \gamma \rightarrow e \gamma^{*} \gamma \rightarrow e t \bar{t}$ and
$e^{-} e^{+}\rightarrow e^{-} \gamma^{*} \gamma^{*} e^{+} \rightarrow e^{-} t \bar{t}e^{+}$, respectively. In conclusion, it is possible that the CLIC may put limits on the electromagnetic dipole moments of the top quark with a sensitivity of the order ${\cal O}(10^{-3}-10^{-2})$ at the $95\%\hspace{0.8mm}C.L.$. We can see from the Figs 3-5, the cross section for the negative values  of the $\hat{a}_v$ are smaller than their positive values. This can easily be seen on sensitivity tables: the bounds for the negative values of the $\hat{a}_v$ for increasing luminosity values do not change much.

It is worthwhile to compare the results obtained here with those of Ref. \cite{Sh} which consider the process
$pp\to p\gamma^* \gamma^*  p \to pt\bar t p$ with the LHC running at $\sqrt{s}=14, 33\hspace{0.8mm}TeV$ and with integrated luminosities
of ${\cal L}=100, 300,3000\hspace{0.8mm}fb^{-1}$. Varying one coupling at a time, they find constraints at $68\%\hspace{0.8mm}C.L.$
of the order ${\cal O}(10^{-2}-10^{-1})$. We also note that, while we do consider three systematic errors in our study, the quoted
sensitivities in Ref. \cite{Sh} do not include theoretical uncertainty. Also, the CLIC sensitivity is even better for our processes than for those reported in Ref. \cite{Sh}.

Finally, in Figs. \ref{Fig.6}-\ref{Fig.8} we show the $95\%$ C.L. contours for anomalous $\hat a_V-\hat a_A$ couplings for the processes $\gamma\gamma\rightarrow t \bar{t}$, $e \gamma \rightarrow e \gamma^{*} \gamma \rightarrow e t \bar{t}$ and $e^{-} e^{+}\rightarrow e^{-} \gamma^{*} \gamma^{*} e^{+} \rightarrow e^{-} t \bar{t}e^{+}$ at the CLIC for various integrated luminosities and center-of-mass energies. Among the three combinations shown in these
figures, we find that the strongest simultaneous limits come from the reaction $\gamma\gamma \to t\bar t$ at the CLIC-$3\hspace{0.8mm}TeV$
and ${\cal L}_{int}=2000\hspace{0.8mm}fb^{-1}$ with the $3\sigma$ level.

\section{Conclusions}

The LHC is expected to provide answers to some fundamental questions of the SM.
However, high precision measurements may not be available due to remnants from the strong interactions of proton-proton collisions.
For this reason, the linear collider with high luminosity and energy is a good choice to
complement and extend the LHC physics program. This collider with high
luminosity and energy is extremely important to examine new physics beyond the SM.
The anomalous $t\bar{t}\gamma$ coupling have very strong energy dependencies since
they are characterized by effective Lagrangians that contrains dimensional-high operators.
Thus, the cross section including the anomalous $t\bar{t}\gamma$ coupling has
a higher energy dependence than the SM cross section.
The anomalous $t\bar{t}\gamma$ coupling can be analyzed through the process
$e^{-}e^{+}\rightarrow t \bar{t}$ at the linear colliders.
This process receives contributions from both anomalous $t\bar{t}\gamma$ and $t\bar{t}Z$ couplings.
However, the processes $\gamma\gamma\rightarrow t \bar{t}$, $e \gamma \rightarrow e \gamma^{*} \gamma \rightarrow e t \bar{t}$ and $e^{-} e^{+}\rightarrow e^{-} \gamma^{*} \gamma^{*} e^{+} \rightarrow e^{-} t \bar{t}e^{+}$ isolate $t\bar{t}\gamma$ coupling which provides the possibility to analyze the $t\bar{t}\gamma$ coupling separately
from the $t\bar{t}Z$ coupling. Any signal which conflicts with the SM predictions would be
convincing evidence for new physics effects in $t\bar{t}\gamma$.

In this paper, we carry out a phenomenological study to investigate the sensitivity of the CLIC to the anomalous $t\bar t \gamma$
coupling through the $\gamma\gamma$, $\gamma\gamma^*$ and $\gamma^*\gamma^*$
collision modes followed by the semileptonic decay of the top pair production. We find that with a center-of-mass energy of CLIC-$1.4\hspace{0.8mm}TeV$, integrated luminosity of ${\cal L}=1500\hspace{0.8mm}fb^{-1}$ and CLIC-$3\hspace{0.8mm}TeV$ and integrated luminosity of ${\cal L}=2000\hspace{0.8mm}fb^{-1}$
with systematic uncertainties of $\delta_{sys}=0, 5, 10\hspace{1mm}\%$ at the $95\%\hspace{1mm}C. L.$, it is possible that the
CLIC may put limits on the electromagnetic dipole moments of the top quark $\hat a_V$ and $\hat a_A$ with a sensitivity of
the order ${\cal O}(10^{-3}-10^{-2})$. In addition, it is noteworthy that our bounds on the dipole moments of the top
quark $\hat a_V$ and $\hat a_A$ at $1\hspace{0.8mm}\sigma$ are predicted to be of the order ${\cal O}(10^{-4}-10^{-3})$,
which is an order of magnitude better than those reported in Refs. \cite{Atwood,Polouse,Choi,Polouse1,Aguilar0,Amjad,Juste,Asner,Abe,Aarons,Brau,Baer}.
Finally, we highlight that the sensitivity with the CLIC data is much stronger than that reported in the literature for the
LHC \cite{Juste} and the ILC \cite{Abe,Aguilar0,Amjad} data.  In conclusion, despite the fact that the LHC prospects are already
strong due to its excellent statistic, the sensitivity of ILC and the CLIC is even stronger.

\begin{table}[!ht]
\caption{Sensitivity on the $\hat a_V$ magnetic moment and the $\hat a_A$ electric dipole moment for the process $\gamma \gamma \to t \bar t $.}
\begin{center}
 \begin{tabular}{cccc}
\hline\hline
\multicolumn{4}{c}{ $\sqrt{s}=1.4\hspace{0.8mm}TeV$,  \hspace{1cm}  $95\%$ C.L.}\\
 \hline
 \cline{1-4} ${\cal L}\hspace{0.8mm}(fb^{-1})$  & \hspace{1.5cm} $\delta_{sys}$ & \hspace{1.5cm} $\hat a_V$   & $\hspace{1.7cm} |\hat a_A|$ \\
\hline
                                           500  &\hspace{1.2cm}  $0\%$   &\hspace{1.2cm} [-0.5170, 0.0034]    & \hspace{1.5cm}   0.0385  \\
                                           500  &\hspace{1.2cm}  $5\%$   &\hspace{1.2cm} [-0.5650, 0.0347]    & \hspace{1.5cm}   0.1286  \\
                                           500  &\hspace{1.2cm}  $10\%$  &\hspace{1.2cm} [-0.6122, 0.0641]    & \hspace{1.5cm}   0.1811  \\
\hline
                                           1000 &\hspace{1.2cm}  $0\%$   &\hspace{1.2cm} [-0.5155, 0.0024]    & \hspace{1.5cm}   0.0324  \\
                                           1000 &\hspace{1.2cm}  $5\%$   &\hspace{1.2cm} [-0.5649, 0.0346]    & \hspace{1.5cm}   0.1285  \\
                                           1000 &\hspace{1.2cm}  $10\%$  &\hspace{1.2cm} [-0.6122, 0.0641]    & \hspace{1.5cm}   0.1811  \\
\hline
                                           1500 &\hspace{1.2cm}  $0\%$   &\hspace{1.2cm} [-0.5149, 0.0020]    & \hspace{1.5cm}   0.0293  \\
                                           1500 &\hspace{1.2cm}  $5\%$   &\hspace{1.2cm} [-0.5648, 0.0346]    & \hspace{1.5cm}   0.1284  \\
                                           1500 &\hspace{1.2cm}  $10\%$  &\hspace{1.2cm} [-0.6121, 0.0640]    & \hspace{1.5cm}   0.1811  \\
\hline\hline
\end{tabular}
\end{center}
\end{table}

\begin{table}[!ht]
\caption{Sensitivity on the $\hat a_V$ magnetic moment and the $\hat a_A$ electric dipole moment for the process $\gamma \gamma \to t \bar t $.}
\begin{center}
 \begin{tabular}{cccc}
\hline\hline
\multicolumn{4}{c}{ $\sqrt{s}=3\hspace{0.8mm}TeV$, \hspace{1cm} $95\%$ C.L.}\\
 \hline
 \cline{1-4}  ${\cal L}\hspace{0.8mm}(fb^{-1})$  & \hspace{1.5cm} $\delta_{sys}$ & \hspace{1.5cm} $\hat a_V$   & $\hspace{1.7cm} |\hat a_A|$ \\
\hline
\hline
  500  &\hspace{1.2cm}  $0\%$   &\hspace{1.2cm} [-0.2225, 0.0040]    & \hspace{1.5cm}   0.0291  \\
  500  &\hspace{1.2cm}  $5\%$   &\hspace{1.2cm} [-0.2564, 0.0331]    & \hspace{1.5cm}   0.0892  \\
  500 &\hspace{1.2cm}   $10\%$  &\hspace{1.2cm} [-0.2870, 0.0585]    & \hspace{1.5cm}   0.1254  \\
\hline
  1000 &\hspace{1.2cm}  $0\%$   &\hspace{1.2cm} [-0.2212, 0.0029]    & \hspace{1.5cm}   0.0245  \\
  1000 &\hspace{1.2cm}  $5\%$   &\hspace{1.2cm} [-0.2563, 0.0330]    & \hspace{1.5cm}   0.0891  \\
 1000 &\hspace{1.2cm}  $10\%$  &\hspace{1.2cm} [-0.2869, 0.0585]    & \hspace{1.5cm}   0.1254  \\
\hline
  1500 &\hspace{1.2cm}  $0\%$   &\hspace{1.2cm} [-0.2206, 0.0024]    & \hspace{1.5cm}   0.0221  \\
  1500 &\hspace{1.2cm}  $5\%$   &\hspace{1.2cm} [-0.2563, 0.0330]    & \hspace{1.5cm}   0.0890  \\
  1500 &\hspace{1.2cm}  $10\%$  &\hspace{1.2cm} [-0.2869, 0.0585]    & \hspace{1.5cm}   0.1254  \\
\hline
  2000 &\hspace{1.2cm}  $0\%$   &\hspace{1.2cm} [-0.2203, 0.0020]    & \hspace{1.5cm}   0.0206  \\
  2000 &\hspace{1.2cm}  $5\%$   &\hspace{1.2cm} [-0.2563, 0.0330]    & \hspace{1.5cm}   0.0879  \\
  2000 &\hspace{1.2cm}  $10\%$  &\hspace{1.2cm} [-0.2869, 0.0585]    & \hspace{1.5cm}   0.1254  \\
\hline\hline
\end{tabular}
\end{center}
\end{table}

\begin{table}[!ht]
\caption{Sensitivity on the $\hat a_V$ magnetic moment and the $\hat a_A$ electric dipole moment for the process $e \gamma \rightarrow e \gamma^{*} \gamma \rightarrow e t \bar{t}$.}
\begin{center}
 \begin{tabular}{cccc}
\hline\hline
\multicolumn{4}{c}{ $\sqrt{s}=1.4\hspace{0.8mm}TeV$,  \hspace{1cm} $95\%$ C.L.}\\
 \hline
 \cline{1-4}  ${\cal L}\hspace{0.8mm}(fb^{-1})$  & \hspace{1.5cm} $\delta_{sys}$ & \hspace{1.5cm} $\hat a_V$   & $\hspace{1.7cm} |\hat a_A|$ \\
\hline
  500  &\hspace{1.2cm}  $0\%$   &\hspace{1.2cm} [-0.7300, 0.0121]    & \hspace{1.5cm}   0.0848  \\
  500  &\hspace{1.2cm}  $5\%$   &\hspace{1.2cm} [-0.7717, 0.0358]    & \hspace{1.5cm}   0.1496  \\
  500 &\hspace{1.2cm}   $10\%$  &\hspace{1.2cm} [-0.8244, 0.0647]    & \hspace{1.5cm}   0.2068  \\
\hline
  1000 &\hspace{1.2cm}  $0\%$   &\hspace{1.2cm} [-0.7241, 0.0086]    & \hspace{1.5cm}   0.0713  \\
  1000 &\hspace{1.2cm}  $5\%$   &\hspace{1.2cm} [-0.7701, 0.0349]    & \hspace{1.5cm}   0.1477  \\
  1000 &\hspace{1.2cm}  $10\%$  &\hspace{1.2cm} [-0.8236, 0.0643]    & \hspace{1.5cm}   0.2061  \\
\hline
  1500 &\hspace{1.2cm}  $0\%$   &\hspace{1.2cm} [-0.7215, 0.0070]    & \hspace{1.5cm}   0.0644  \\
  1500 &\hspace{1.2cm}  $5\%$   &\hspace{1.2cm} [-0.7697, 0.0346]    & \hspace{1.5cm}   0.1470  \\
  1500 &\hspace{1.2cm}  $10\%$  &\hspace{1.2cm} [-0.8234, 0.0641]    & \hspace{1.5cm}   0.2059  \\
\hline\hline
\end{tabular}
\end{center}
\end{table}

\begin{table}[!ht]
\caption{Sensitivity on the $\hat a_V$ magnetic moment and the $\hat a_A$ electric dipole moment for the process $e \gamma \rightarrow e \gamma^{*} \gamma \rightarrow e t \bar{t}$.}
\begin{center}
 \begin{tabular}{cccc}
\hline\hline
\multicolumn{4}{c}{ $\sqrt{s}=3\hspace{0.8mm}TeV$, \hspace{1cm}   $95\%$ C.L.}\\
 \hline
 \cline{1-4}  ${\cal L}\hspace{0.8mm}(fb^{-1})$  & \hspace{1.5cm} $\delta_{sys}$ & \hspace{1.5cm} $\hat a_V$   & $\hspace{1.7cm} |\hat a_A|$ \\
\hline
  500  &\hspace{1.2cm}  $0\%$   &\hspace{1.2cm} [-0.4626, 0.0088]    & \hspace{1.5cm}   0.0610  \\
  500  &\hspace{1.2cm}  $5\%$   &\hspace{1.2cm} [-0.4961, 0.0034]    & \hspace{1.5cm}   0.1249  \\
  500  &\hspace{1.2cm}   $10\%$  &\hspace{1.2cm} [-0.5333, 0.0630]    & \hspace{1.5cm}   0.1740  \\
\hline
  1000 &\hspace{1.2cm}  $0\%$   &\hspace{1.2cm} [-0.4593, 0.0063]    & \hspace{1.5cm}   0.0512  \\
  1000 &\hspace{1.2cm}  $5\%$   &\hspace{1.2cm} [-0.4955, 0.0343]    & \hspace{1.5cm}   0.1240  \\
  1000 &\hspace{1.2cm}  $10\%$  &\hspace{1.2cm} [-0.5332, 0.0629]    & \hspace{1.5cm}   0.1737  \\
\hline
  1500 &\hspace{1.2cm}  $0\%$   &\hspace{1.2cm} [-0.4579, 0.0052]    & \hspace{1.5cm}   0.0463  \\
  1500 &\hspace{1.2cm}  $5\%$   &\hspace{1.2cm} [-0.4953, 0.0342]    & \hspace{1.5cm}   0.1237  \\
  1500 &\hspace{1.2cm}  $10\%$  &\hspace{1.2cm} [-0.5331, 0.0628]    & \hspace{1.5cm}   0.1736  \\
\hline
 2000 &\hspace{1.2cm}  $0\%$   &\hspace{1.2cm} [-0.4570, 0.0045]    & \hspace{1.5cm}   0.0431  \\
 2000 &\hspace{1.2cm}  $5\%$   &\hspace{1.2cm} [-0.4952, 0.0341]    & \hspace{1.5cm}   0.1236  \\
 2000 &\hspace{1.2cm}  $10\%$  &\hspace{1.2cm} [-0.5330, 0.0627]    & \hspace{1.5cm}   0.1735  \\
\hline\hline
\end{tabular}
\end{center}
\end{table}

\begin{table}[!ht]
\caption{Sensitivity on the $\hat a_V$ magnetic moment and the $\hat a_A$ electric dipole moment for the process $e^{-} e^{+}\rightarrow e^{-} \gamma^{*} \gamma^{*} e^{+} \rightarrow e^{-} t \bar{t}e^{+}$.}
\begin{center}
 \begin{tabular}{cccc}
\hline\hline
\multicolumn{4}{c}{ $\sqrt{s}=1.4\hspace{0.8mm}TeV$,  \hspace{1cm}  $95\%$ C.L.}\\
 \hline
 \cline{1-4}  ${\cal L}\hspace{0.8mm}(fb^{-1})$  & \hspace{1.5cm} $\delta_{sys}$ & \hspace{1.5cm} $\hat a_V$   & $\hspace{1.7cm} |\hat a_A|$ \\
\hline
  500  &\hspace{1.2cm}  $0\%$   &\hspace{1.2cm} [-0.9123, 0.0490]    & \hspace{1.5cm}   0.1878  \\
  500  &\hspace{1.2cm}  $5\%$   &\hspace{1.2cm} [-0.9298, 0.0580]    & \hspace{1.5cm}   0.2057  \\
  500 &\hspace{1.2cm}   $10\%$  &\hspace{1.2cm} [-0.9690, 0.0774]    & \hspace{1.5cm}   0.2415  \\
\hline
  1000 &\hspace{1.2cm}  $0\%$   &\hspace{1.2cm} [-0.8859, 0.0357]    & \hspace{1.5cm}   0.1581  \\
  1000 &\hspace{1.2cm}  $5\%$   &\hspace{1.2cm} [-0.9090, 0.0478]    & \hspace{1.5cm}   0.1850  \\
  1000 &\hspace{1.2cm}  $10\%$  &\hspace{1.2cm} [-0.9558, 0.0709]    & \hspace{1.5cm}   0.2299  \\
\hline
  1500 &\hspace{1.2cm}  $0\%$   &\hspace{1.2cm} [-0.8739, 0.0295]    & \hspace{1.5cm}   0.1429  \\
 1500 &\hspace{1.2cm}  $5\%$   &\hspace{1.2cm} [-0.9014, 0.0437]    & \hspace{1.5cm}   0.1762  \\
 1500 &\hspace{1.2cm}  $10\%$  &\hspace{1.2cm} [-0.9510, 0.0686]    & \hspace{1.5cm}   0.2256  \\
\hline\hline
\end{tabular}
\end{center}
\end{table}

\begin{table}[!ht]
\caption{Sensitivity on the $\hat a_V$ magnetic moment and the $\hat a_A$ electric dipole moment for the process $e^{-} e^{+}\rightarrow e^{-} \gamma^{*} \gamma^{*} e^{+} \rightarrow e^{-} t \bar{t}e^{+}$.}
\begin{center}
 \begin{tabular}{cccc}
\hline\hline
\multicolumn{4}{c}{ $\sqrt{s}=3\hspace{0.8mm}TeV$, \hspace{1cm}   $95\%$ C.L.}\\
 \hline
 \cline{1-4}  ${\cal L}\hspace{0.8mm}(fb^{-1})$  & \hspace{1.5cm} $\delta_{sys}$ & \hspace{1.5cm} $\hat a_V$   & $\hspace{1.7cm} |\hat a_A|$ \\
\hline
  500  &\hspace{1.2cm}  $0\%$   &\hspace{1.2cm} [-0.6212, 0.0291]    & \hspace{1.5cm}   0.1259  \\
  500  &\hspace{1.2cm}  $5\%$   &\hspace{1.2cm} [-0.6417, 0.0434]    & \hspace{1.5cm}   0.1561  \\
  500 &\hspace{1.2cm}   $10\%$  &\hspace{1.2cm} [-0.6773, 0.0679]    & \hspace{1.5cm}   0.2000  \\
\hline
  1000 &\hspace{1.2cm}  $0\%$   &\hspace{1.2cm} [-0.6097, 0.0210]    & \hspace{1.5cm}   0.1059  \\
  1000 &\hspace{1.2cm}  $5\%$   &\hspace{1.2cm} [-0.6353, 0.0390]    & \hspace{1.5cm}   0.1472  \\
  1000 &\hspace{1.2cm}  $10\%$  &\hspace{1.2cm} [-0.6739, 0.0656]    & \hspace{1.5cm}   0.1962  \\
\hline
  1500 &\hspace{1.2cm}  $0\%$   &\hspace{1.2cm} [-0.6044, 0.0173]    & \hspace{1.5cm}   0.0957  \\
  1500 &\hspace{1.2cm}  $5\%$   &\hspace{1.2cm} [-0.6330, 0.0374]    & \hspace{1.5cm}   0.1439  \\
  1500 &\hspace{1.2cm}  $10\%$  &\hspace{1.2cm} [-0.6728, 0.0648]    & \hspace{1.5cm}   0.1948  \\
\hline
  2000 &\hspace{1.2cm}  $0\%$   &\hspace{1.2cm} [-0.6013, 0.0151]    & \hspace{1.5cm}   0.0890  \\
  2000 &\hspace{1.2cm}  $5\%$   &\hspace{1.2cm} [-0.6317, 0.0365]    & \hspace{1.5cm}   0.1420  \\
  2000 &\hspace{1.2cm}  $10\%$  &\hspace{1.2cm} [-0.6721, 0.0644]    & \hspace{1.5cm}   0.1942  \\
\hline\hline
\end{tabular}
\end{center}
\end{table}

\newpage

\begin{center}
{\bf Acknowledgments}
\end{center}

A. G. R. acknowledges support from CONACyT, SNI and PROFOCIE (M\'exico).


\pagebreak

\begin{figure}[t]
\centerline{\scalebox{0.75}{\includegraphics{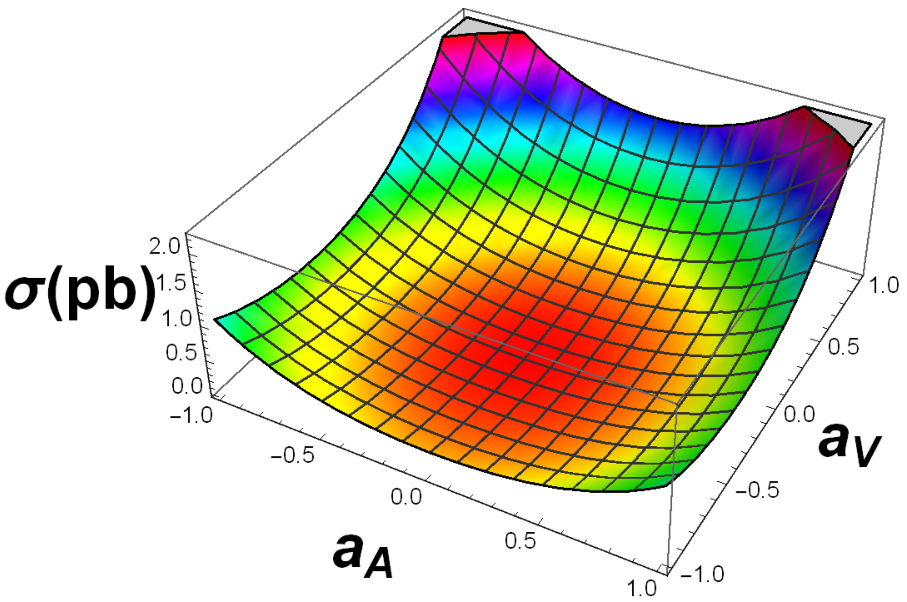}}}
\caption{ \label{fig:gamma2} Same as in Fig. 5, but for center-of-mass energy of
$\sqrt{s}=3$ .}
\label{Fig.6}
\end{figure}

\begin{figure}[t]
\centerline{\scalebox{0.75}{\includegraphics{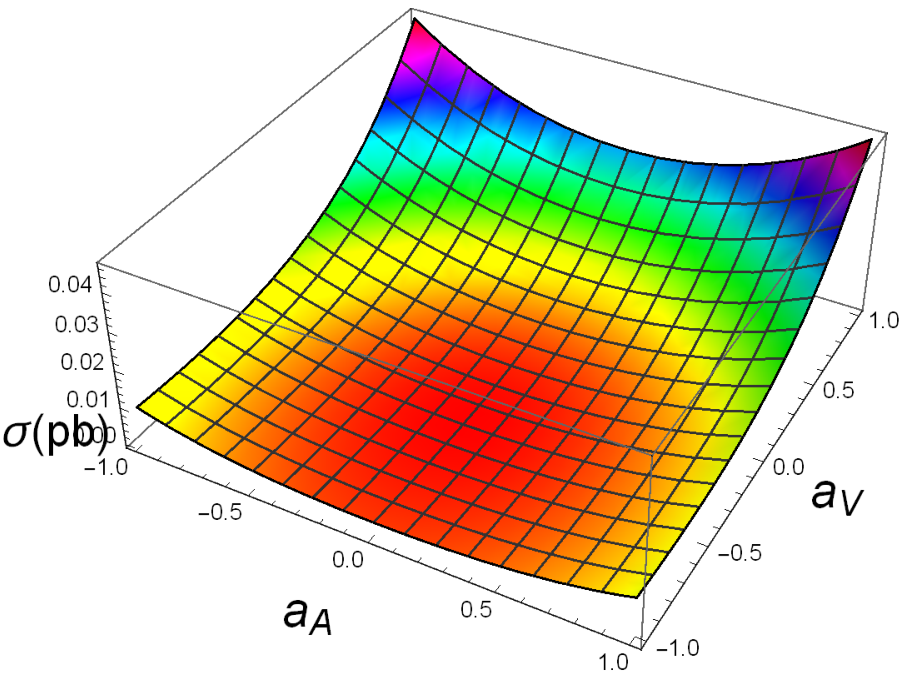}}}
\caption{ \label{fig:gamma1x} The total cross sections of the process
$e^+e^- \to e^+\gamma^* \gamma^* e^- \to e^+ t \bar t e^-$ as a function of $\hat a_V$ and $\hat a_A$ for center-of-mass energy of
$\sqrt{s}=1.4$\hspace{0.8mm}$TeV$.}
\label{Fig.7}
\end{figure}

\begin{figure}[t]
\centerline{\scalebox{0.75}{\includegraphics{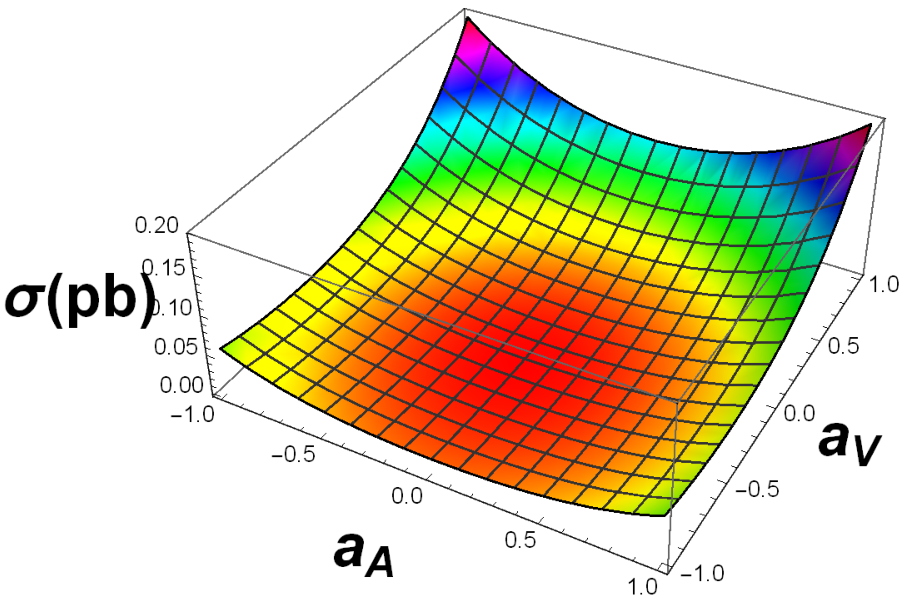}}}
\caption{ \label{fig:gamma2x} Same as in Fig. 7, but for center-of-mass energy of $\sqrt{s}=3$ .}
\label{Fig.8}
\end{figure}

\begin{figure}[t]
\centerline{\scalebox{0.77}{\includegraphics{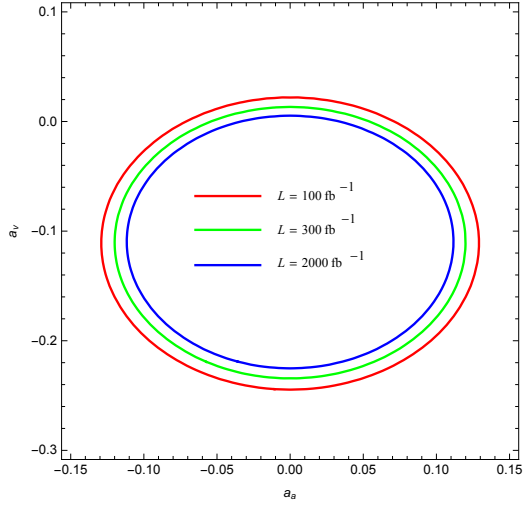}}}
\caption{ \label{fig:gamma15}  Bounds contours at the $68\% \hspace{1mm}C.L.$ in the
$\hat a_V-\hat a_A$ plane for the process $\gamma \gamma \to t \bar t $ with the $\delta _{sys}=0\%$ and for center-of-mass energy of $\sqrt{s}=3$.}
\label{Fig.6}
\end{figure}

\begin{figure}[t]
\centerline{\scalebox{0.76}{\includegraphics{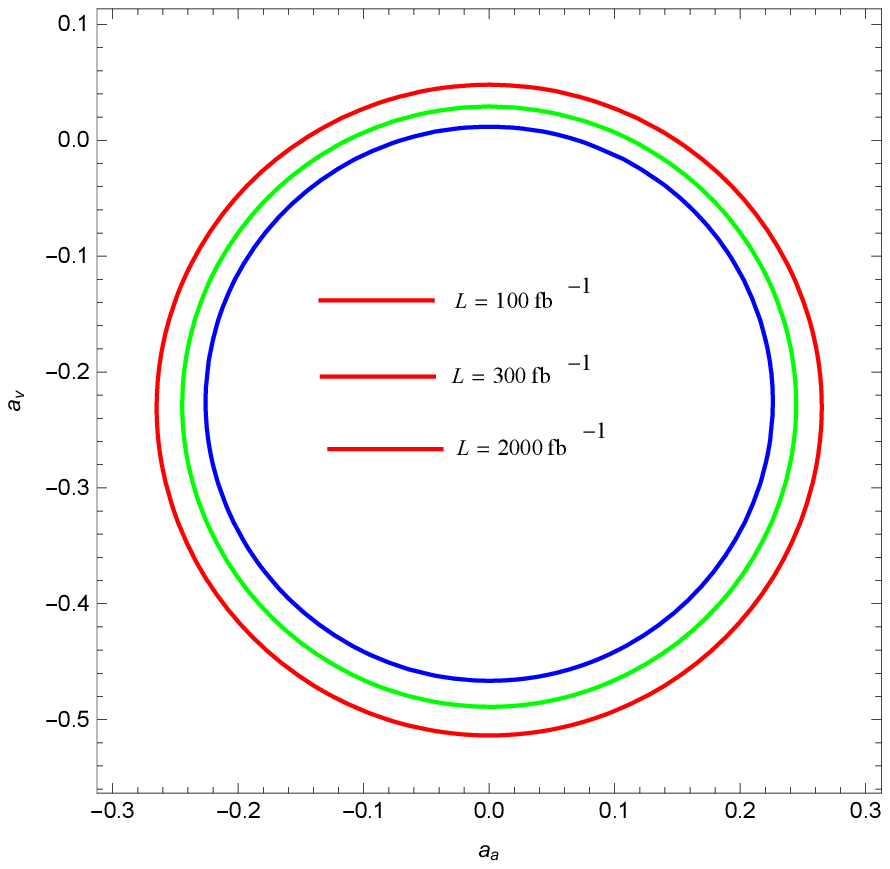}}}
\caption{ \label{fig:gamma6} Bounds contours at the $68\% \hspace{1mm}C.L.$ in the
$\hat a_V-\hat a_A$ plane for $e^+e^- \to e^+\gamma\gamma^* e^- \to e^+t \bar t e^-$ with the $\delta _{sys}=0\%$ and for center-of-mass energy of $\sqrt{s}=3$.}
\label{Fig.7}
\end{figure}

\begin{figure}[t]
\centerline{\scalebox{0.76}{\includegraphics{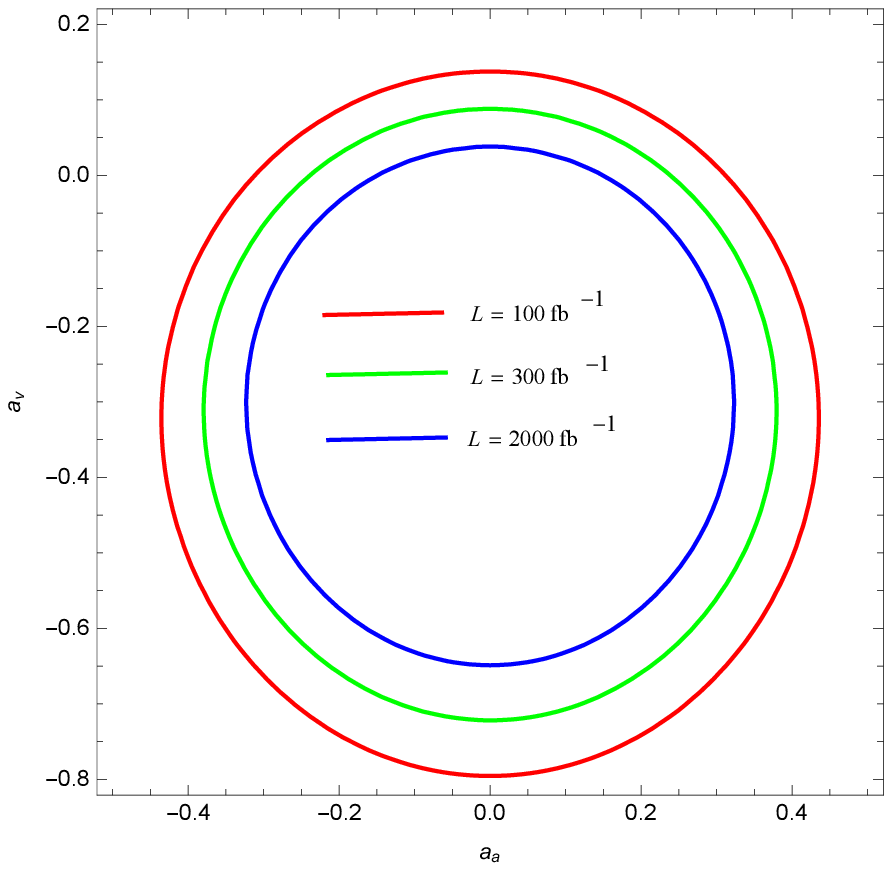}}}
\caption{ \label{fig:gamma6x} Bounds contours at the $68\% \hspace{1mm}C.L.$ in the
$\hat a_V-\hat a_A$ plane for $e^+e^- \to e^+\gamma^*\gamma^* e^- \to e^+t \bar t e^-$ with the $\delta _{sys}=0\%$ and for center-of-mass energy of $\sqrt{s}=3$.}
\label{Fig.8}
\end{figure}

\end{document}